\documentclass[aps,prd,twocolumn,groupedaddress,showpacs,nofootinbib]{revtex4}
\usepackage{graphicx}

\begin{document}

\title{Dark energy exponential potential models as curvature quintessence}

\author{S. Capozziello$^1$, V.F. Cardone$^2$, E. Piedipalumbo$^1$, C. Rubano$^1$}
\thanks{Corresponding author\,: V.F. Cardone, {\tt winny@na.infn.it}.}

\affiliation{$^1$Dipartimento di Scienze Fisiche, Univ. di Napoli "Federico II", and INFN, Sez. di Napoli, Compl. Univ. di Monte S. Angelo, Ed. N, via Cinthia, 80121 - Napoli, Italy \\ $^2$Dipartimento di Fisica "E.R. Caianiello",  Univ. di Salerno, and INFN, Sez. di Napoli, Gruppo Coll. di Salerno, via S. Allende, 84081 - Baronissi (Salerno), Italy}

\begin{abstract}

It has been recently shown that, under some general conditions, it is always possible to find a fourth order gravity theory capable of reproducing the same dynamics of a given dark energy model. Here, we discuss this approach for a dark energy model with a scalar field evolving under the action of an exponential potential. In absence of matter, such a potential can be recovered from a fourth order theory via a conformal transformation. Including the matter term, the function $f(R)$ entering the generalized gravity Lagrangian can be reconstructed according to the dark energy model.

\end{abstract}


\maketitle

\section{Introduction}

The era of {\it precision cosmology} has opened the way to a new
picture of the universe depicted as a spatially flat manifold with a
subcritical matter content undergoing a phase of accelerated
expansion. Although totally unexpected only few years ago, this
scenario is nowadays fully accepted as the standard one, thanks to
the impressive amount of astrophysical evidences on different
scales. The anisotropy spectrum of cosmic microwave background
radiation \cite{CMBR,WMAP}, the SNeIa Hubble diagram \cite{SNeIa},
the large scale structure \cite{LSS} and the matter power spectrum
determined by the Ly$\alpha$ forest data \cite{Lyalpha} represent
observational cornerstones that put on firm grounds the picture of
this kind of universe model.

While observationally well founded, this scenario is plagued by the
unsolved puzzle of {\it dark energy}, the negative pressure fluid
which seems to be necessary to both close the universe and drive its
accelerated expansion. Investigating the nature and the properties
of this mysterious component is one of the most exciting problems of
modern cosmology, with proposals ranging from the classical
cosmological constant \cite{Lambda}, to scalar field quintessence
\cite{QuintRev} and unified dark energy models in which a single
fluid with an exotic equation of state accounts for both dark energy
and dark matter \cite{Chaplygin,tachyon,Hobbit}.

Notwithstanding the strong efforts made to solve this puzzle, none
of the proposed explanations is fully satisfactory and free of
problems. This disturbing situation has motivated much interest
towards radically different approaches to the problem of dark
energy. After all, dark energy is usually introduced only as a tool
to explain the cosmic acceleration, while there is no direct
evidence of its existence as a separate fluid. Motivated by this
consideration, it has been suggested that cosmic speed up could be
an evidence for the need of {\it new physics} rather than a new
fluid. Much interest has then been devoted, for example, to models
according to which standard matter is the only physical ingredient,
while the Friedmann equations are modified as in the Cardassian
expansion \cite{Cardassian}, the Dvali\,-\,Gabadadze\,-\,Porrati
scenario \cite{DGP} or the improved renormalization group approach
\cite{Alfio}.

In this same framework, higher order theories of gravity also
represent a valid alternative to the dark energy approach. In these
models, also referred to as {\it curvature quintessence}
\cite{capozcurv}, the gravity Lagrangian is generalized by replacing
the Ricci scalar curvature $R$ with a generic function $f(R)$, so
that an effective dark energy\,-\,like fluid appears in the
Friedmann equations and drives the accelerated expansion. Reasons
for considering non linear curvature terms in the gravity Lagrangian
can be found, for instance, in quantum effects on curved spacetimes
or in certain low energy limits of string/M theories. Moreover,
early time inflationary behaviours can be naturally recovered from
nonlinear gravity Lagrangians thus further motivating to study
possible effects of such terms on the late time universe expansion.
Different models of this kind have been explored and tested against
observations also considering the two possible formulations that are
obtained adopting the metric
\cite{capozcurv,review,noicurv,MetricRn} or the Palatini
\cite{PalRn,lnR,Allemandi} formulations.

Recently, it has been shown \cite{CCT} that, under some quite
general conditions, the dynamics (i.e., the scale factor and the
expansion rate) of a given dark energy model may be, in principle,
reproduced as a byproduct of a $f(R)$ theory of gravity\footnote{See
also \cite{danish} for some comments on the uniqueness of the
reconstruction in the matter only case.}. The method developed in
\cite{CCT} makes it possible to derive the $f(R)$ theory which
reproduces a given $H(z)$ and represents therefore a sort of {\it
bridge} between two physically different scenarios. Here, we apply
the method to a quintessence model where the evolution of the scalar
field is determined by an exponential potential \cite{RS01,moki}.
Since this model has been shown to fit well a large set of
observational constraints \cite{pavlov,mauro,ester}, we are thus
able to recover a $f(R)$ theory that reproduces the same data set
with the same accuracy, so being motivated {\it a posteriori} and
showing that the two approaches are phenomenologically equivalent.

The plan of the paper is as follows. In Sects.\,II and III, we shortly review the main features of the dark energy model and of the curvature quintessence scenario, respectively. In Sect.\,III, the procedure adopted to recover the $f(R)$ theory representing the higher order counterpart of a given dark energy model is presented. In Sect.\,IV, we use conformal transformations to show the equivalence between a class of $f(R)$ theories and a given scalar field model with exponential potential. Since this equivalence only holds in absence of matter, Sect.\,IV presents the results obtained applying the method outlined in Sect.\,III to recover the $f(R)$ Lagrangian corresponding to the exponential potential model in presence of the matter term. Some caveats related to our technique are commented upon in Sect.\,V, where we also summarize and conclude.

\section{The exponential potential dark energy model}

A standard recipe to explain the observed cosmic acceleration consists in considering a scalar field $\varphi$ minimally coupled to gravity yielding a homogenously distributed fluid with a negative equation of state. The energy density and pressure for such a fluid read\footnote{We use here natural units such that $8 \pi G_N = c = 1$.}

\begin{equation}
\left \{
\begin{array}{l}
\rho_{\varphi} = \frac{1}{2} \dot{\varphi}^2 + V(\varphi) \\
~ \nonumber \\
p_{\varphi} = \frac{1}{2} \dot{\varphi}^2 - V(\varphi)
\end{array}
\right .
\label{eq: phi}
\end{equation}
with an overdot denoting derivation with respect to cosmic time $t$ and $V(\varphi)$ the self interaction potential. As it is clear, the choice of $V(\varphi)$ plays a key role in determining the dynamics of the universe.

As an interesting application, we consider the simple exponential potential \cite{RS01,moki,pavlov,mauro,ester}\,:

\begin{equation}
V(\varphi) \propto \exp{\left ( - \sqrt{\frac{3}{2}} \varphi \right )}
\label{eq: exppot}
\end{equation}
which presents the nice feature that the corresponding Friedmann equations may be analytically solved. We refer the interested reader to \cite{RS01}, while here we only report the quantities we will need in the following. In a spatially flat universe, the scale factor $a$ and the Hubble parameter $H = \dot{a}/a$ read \cite{RS01,ester}\,:

\begin{equation}
a^3 = \left [ (3 {\cal{H}}_0 - 2) \tau^2 + 4 - 3 {\cal{H}}_0 \right ] \frac{\tau^2}{2} \ ,
\label{eq: atsingle}
\end{equation}

\begin{equation}
H = \frac{2 \left [ 2 \left ( 3 {\cal{H}}_0 - 2 \right ) \tau^2 + 4 - 3 {\cal{H}}_0 \right ]}{3 \tau \left [ 2 \left ( 3 {\cal{H}}_0 - 2 \right ) \tau^2 + 4 - 3 {\cal{H}}_0 \right ]}
\label{eq: htsingle}
\end{equation}
with $\tau = t/t_0$, $t_0$ the present age of the universe and ${\cal{H}}_0$ the Hubble constant in units of $t_0$ and hereon the subscript 0 will refer to quantities evaluated today. Taking $t_0$ as unit of time, the model is fully characterized by the parameter ${\cal{H}}_0$. In particular, the present day matter density parameter turns out to be \cite{ester}\,:

\begin{equation}
\Omega_M = \frac{2 (4 - 3 {\cal{H}}_0)}{9 {\cal{H}}_0^2} \ ,
\label{eq: omsingle}
\end{equation}
which, for the best fit value ${\cal{H}}_0 = 0.97$ \cite{ester}, gives $\Omega_M \simeq 0.26$, in very good agreement with recent estimates.

\section{Curvature quintessence}

Much interest has been recently devoted to the so called curvature quintessence according to which the universe is filled by pressureless dust matter only and the acceleration is the result of the modified Friedmann equations obtained by replacing the Ricci scalar curvature $R$ with a generic function $f(R)$ in the gravity Lagrangian. The Friedmann equations therefore read \cite{capozcurv,review}\,:

\begin{equation}
H^2 = \frac{1}{3} \left [ \frac{\rho_m}{f'(R)} + \rho_{curv} \right ]  \ ,
\label{eq: hfr}
\end{equation}

\begin{equation}
2 \frac{\ddot{a}}{a} + H^2 = - w_{curv} \rho_{curv} \ ,
\label{eq: fried2}
\end{equation}
where the prime denotes derivative with respect to $R$, $\rho_{curv}$ is the energy density of an {\it effective curvature fluid} given as\,:

\begin{equation}
\rho_{curv} = \frac{1}{f'(R)} \left \{ \frac{1}{2} \left [ f(R)  - R f'(R) \right ] - 3 H \dot{R} f''(R) \right \} \ ,
\label{eq: rhocurv}
\end{equation}
and the barotropic factor of the curvature fluid is\,:

\begin{equation}
w_{curv} = -1 + \frac{\ddot{R} f''(R) + \dot{R} \left [ \dot{R} f'''(R) - H f''(R) \right ]} {\left [ f(R) - R f'(R) \right ]/2 - 3 H \dot{R} f''(R)}
\label{eq: wcurv}
\end{equation}
Assuming that there is no interaction between matter and curvature terms, the continuity equation for $\rho_{curv}$ reads \cite{CCT}\,:

\begin{equation}
\dot{\rho}_{curv} + 3 H (1 + w_{curv}) \rho_{curv}  = \frac{3 H_0^2 \Omega_M \dot{R} f''(R)}{\left [ f'(R) \right ]^2}  a^{-3}
\label{eq: curvcons}
\end{equation}
which is identically satisfied as can be easily shown using Eq.(\ref{eq: hfr}) and expressing the scalar curvature $R$ as function of the Hubble parameter\,:
\begin{equation}
R = - 6 \left ( \dot{H} + 2 H^2 \right ) \ .
\label{eq: rvsh}
\end{equation}
Combining Eqs.(\ref{eq: hfr}) with Eq.(\ref{eq: fried2}) and using the definition of $H$, one finally gets the following {\it master equation} for the Hubble parameter \cite{CCT}\,:

\begin{eqnarray}
\dot{H} & = & -\frac{1}{2 f'(R)} \left \{ 3 H_0^2 \Omega_M a^{-3} + \ddot{R} f''(R)+ \right . \nonumber \\
~ & ~ & \left . + \dot{R} \left [ \dot{R} f'''(R) - H f''(R) \right ] \right \} \ .
\label{eq: presingleeq}
\end{eqnarray}
Inserting Eq.(\ref{eq: rvsh}) into Eq.(\ref{eq: presingleeq}), one ends with a fourth order nonlinear differential equation for the scale factor $a(t)$ that cannot be analitically solved also for the simplest cases (for instance, $f(R) \propto R^n$ unless dust matter contribution is discarded). Moreover, although technically feasible, a numerical solution of Eq.(\ref{eq: presingleeq}) is plagued by the large uncertainties on the boundary conditions (i.e., the present day values of the scale factor and its derivatives up to the third order) that have to be set to find out $a(t)$ by solving Eq.(\ref{eq: presingleeq}).

Given these mathematical difficulties, a different approach has been proposed in \cite{CCT} (hereafter CCT) where Eq.(\ref{eq: presingleeq}) is considered as a way to determine $f(R)$ rather than $a(t)$. Rearranging the different terms suitably, CCT obtained a linear third order differential equation for $f$ in terms of the redshift $z = 1/a - 1$ (having set $a_0 = 1$) that can be easily solved numerically for a given $H(z)$. By this method, it is then possible to find out which $f(R)$ theory reproduces the same dynamics of a given dark energy model, thus showing a formal equivalence between these two radically different approaches.

CCT developed the method using the redshift $z$ as integration variable since it is common to have an analytical expression for the Hubble parameter as function of $z$ rather than $t$. However, this is not the case for the models considered here, so that we rewrite the main formulae in CCT using $t$ as integration variable. The equation determining $f(t)$ is then\footnote{With an abuse of notation, we write $f(t)$ rather than $f[R(t)]$.}\,:

\begin{equation}
{\cal{H}}_3(t) \frac{d^3f}{dt^3} + {\cal{H}}_2(t) \frac{d^2f}{dt^2} + {\cal{H}}_1(t) \frac{df}{dt} = - 3 H_0^2 \Omega_M \dot{R}^2 a^{-3}(t) \ ,
\label{eq: singleeq}
\end{equation}
with\,:
\begin{equation}
{\cal{H}}_1 = 2 \dot{H} \dot{R} + H \ddot{R} + 2 \ddot{R}^2 \dot{R}^{-1} - \dddot{R} \ ,
\end{equation}

\begin{equation}
{\cal{H}}_2 = - \left ( 2 \ddot{R} + H \dot{R} \right ) \ ,
\end{equation}

\begin{equation}
{\cal{H}}_3 = \dot{R} \ .
\end{equation}
where $R$ is given by Eq.(\ref{eq: rvsh}). In order to integrate Eq.(\ref{eq: singleeq}), we need to specify boundary conditions that are more conveniently assigned at the present time. We slightly generalize here the discussion presented in CCT. First, let us remember that, in a fourth order theory, we may define an {\it effective gravitational constant} as $G_{eff} = G_N/f'(R)$, with $G_N$ the usual Newtonian gravitational constant. Its rate of variation will be given as\,:

\begin{equation}
\frac{\dot{G}_{eff}}{G_{eff}} = - \frac{1}{t_0} \frac{f''(R)}{f'(R)} \frac{dR}{d\tau} \ .
\label{eq: geffvar}
\end{equation}
It is quite natural to assume that the effective and the Newtonian gravitational couplings take the same values today, so that we get the condition\,:

\begin{equation}
f'(R_0) = 1 \ . \label{eq: in1}
\end{equation}
Evaluating Eq.(\ref{eq: geffvar}) for $t = t_0$ (i.e., $\tau = 1$), we may determine $f''(R_0)$ provided an estimate of $(\dot{G}_{eff}/G_{eff})_{t= t_0}$ is given. Since in our theory $G_N$ is constant, we may assume that the measurements of the variation of $G_N$ \cite{uzan} actually refers to $G_{eff}$ and use these results to get an estimate of $(\dot{G}_{eff}/G_{eff})_{t= t_0}$. We thus take as our boundary condition\,:

\begin{equation}
f''(R_0) = - t_0 \left ( \frac{\dot{G}_{eff}}{G_{eff}} \right )_{obs} \left ( \frac{dR}{d\tau} \right )^{-1} \ ,
\label{eq: in2}
\end{equation}
having used Eq.(\ref{eq: in1}). Finally, inserting Eqs.(\ref{eq: in1}) and (\ref{eq: in2}) into Eq.(\ref{eq: rhocurv}) and then in (\ref{eq: hfr}) evaluated today, we get\,:

\begin{eqnarray}
f(R_0) & = & 6 H_0^2 \left ( 1 - \Omega_M + \frac{R_0}{6 H_0^2} \right ) f'(R_0) \nonumber \\ ~ & + & 6 H_0 \left ( \frac{dR}{dt}
\right )_{t = t_0} f''(R_0) \ .
\label{eq: in0}
\end{eqnarray}
It is then straightforward to show that the following boundary conditions descend from Eqs.(\ref{eq: in2})\,-\,(\ref{eq: in0})\,:

\begin{equation}
\left ( \frac{df}{dt} \right )_{t = t_0} = \left ( \frac{dR}{dt} \right )_{t = t_0} f'(R_0) \ ,
\label{eq: in1bis}
\end{equation}

\begin{equation}
\left ( \frac{d^2f}{dt^2} \right )_{t = t_0} = \left ( \frac{dR}{dt} \right )_{t = t_0}^2 f''(R_0) + \left ( \frac{d^2R}{dt^2} \right )_{t = t_0} f'(R_0)
\label{eq: in2bis}
\end{equation}
that have to be used, together with Eq.(\ref{eq: in0}), to numerically solve Eq.(\ref{eq: singleeq}). Combining the solution thus obtained for $f(t)$ with $R(t)$ evaluated through Eq.(\ref{eq: rvsh}), one finally finds $f(R)$ thus recovering the higher order theory that mimicks the assigned dark energy model. We refer the reader to \cite{CCT} for some interesting examples.

\section{Conformal equivalence}

The procedure described above may be easily implemented in the case of the exponential potential model presented in Sect.\,II. Neglecting the matter term (i.e., setting $\Omega_M = 0$), it is however possible to show that there exists a more intimate relation between $f(R)$ theories and a scalar field evolving under the action of an exponential potential.

To this aim, let us remember that, in general, given a $f(R)$ theory in the Jordan frame, the conformal transformation

\begin{displaymath}
g_{\mu \nu} \rightarrow \tilde{g}_{\mu \nu} = \Omega^2 g_{\mu \nu} \ ,
\end{displaymath}
with $\Omega \equiv \exp{(\omega)} = \sqrt{f'(R)}$, leads, in the Einstein frame, to a minimally coupled scalar field theory whose self interaction potential is given by \cite{confpot}\,:

\begin{equation}
V(\tilde{\varphi}) = \frac{f(R) - R f'(R)}{2 \left [ f'(R) \right ]^2 } \ ,
\label{eq: potconf}
\end{equation}
where tilted quantities refer to the Einstein frame. Let us now consider the particular choice\,:

\begin{equation}
f(R) \equiv R + \frac{\alpha}{R}
\label{eq: fconf}
\end{equation}
and recall that the conformal factor may also be rescaled as $\omega = K \tilde{\varphi}$ with $K$ an (up to now) arbitrary constant \cite{ACCF}. Imposing $f'(R) = \exp{(\omega)}$ and solving with respect to $R$, the potential (\ref{eq: potconf}) may be rewritten as\,:

\begin{equation}
V(\varphi) = \frac{\sqrt{| \alpha |}}{2} \exp{(-4 K \varphi)} \sqrt{\exp{(2 K \varphi)} - 1} \ ,
\label{eq: potconfbis}
\end{equation}
where we have assumed $\alpha < 0$ and hereon we will use nontilted quantities for simplicity. Choosing $K = 1/6$ and considering the limit $\exp{(2 K \varphi)} >> 1$, the potential (\ref{eq: potconfbis}) asymptotically behaves as\,:

\begin{displaymath}
V(\varphi) \simeq \exp{\left ( - \sqrt{\frac{3}{2}} \varphi \right )} \ ,
\end{displaymath}
which is just the exponential potential model we are considering in Eq.(\ref{eq: exppot}).

On the other hand, starting directly with $f'(R) = \exp{(2 K \varphi)}$, the potential in Eq.(\ref{eq: potconfbis}) is soon written as

\begin{equation}
V = \frac{f - R \exp{(2 K \varphi)}}{2 \exp{(4 K \varphi)}} \ ,
\label{46}
\end{equation}
so that

\begin{equation}
\frac{dV}{d\varphi} = K \times \frac{R f'(R) - 2 f(R)}{\left [ f'(R) \right ]^2} \ ,
\end{equation}
having used $df/d\varphi = (df/dR)(dR/d\varphi) = f'(R) dR/d\varphi$. Now, it can be easily shown that posing

\begin{equation}
\frac{dV}{d\varphi} = - \lambda V \ ,
\end{equation}
with $\lambda$ an arbitrary constant, leads to $f = f_{0}R^{n}$, where $f_{0}$ is an integration constant and $n = (4K - \lambda)/(2K - \lambda)$. Therefore, starting now from

\begin{equation}
f(R) \equiv R^{n} \ ,
\label{49}
\end{equation}
the potential in Eq.(\ref{eq: potconfbis}) becomes

\begin{displaymath}
V = V(R) = \frac{1 - n}{2n^{2}}R^{2 - n} \Rightarrow
\end{displaymath}

\begin{equation}
V = V(\varphi) = \frac{1 - n}{2}|n|^{\frac{n}{1 - n}}\exp(-\lambda \varphi) \,,
\label{50}
\end{equation}
with $\lambda = 2K(n - 2)/(n - 1)$.

Assuming $K \equiv \sqrt{1/6}$ and $\lambda \equiv \sqrt{3/2}$ gives

\begin{equation}
n = -1 \,,
\label{51}
\end{equation}
so that we get the nice result that a higher order theory of gravity (without matter, anyway) with

\begin{equation}
f(R) \equiv \frac{1}{R}
\label{52}
\end{equation}
is conformally transformed into a usual minimally coupled general relativistic theory with a massive scalar field $\varphi$ only, whose potential is

\begin{equation}
V(\varphi) = \exp \left( -\sqrt{\frac{3}{2}}\varphi \right)
\,.\label{53}
\end{equation}
By means of conformal transformations, it is then possible to find out an intimate link between two different scenarios, namely a particularly simple class of $f(R)$ theories and a minimally coupled scalar field with specific exponential potential. It is worth stressing, however, that the above results only hold in the absence of matter. While it is still possible to work out a conformal transformation also in presence of matter, the self interaction potential may be no more expressed as in Eq.(\ref{eq: potconf}). Although such a problem may be in principle escaped by resorting to numerical methods, it is nevertheless problematic to assess the meaning of conformal transformations, since there is a still open debate regarding the {\it physical} equivalence of two conformally (and hence {\it mathematically}) equivalent theories (see \cite{ACCF} and references therein).

\section{$f(R)$ and exponential potential}

The dark energy model presented in Sect.\,II has nice features both from the mathematical and the physical points of view. Indeed, the main dynamical quantities (such as the scale factor and the expansion rate) may be written analytically, while it very well matches the SNeIa Hubble diagram, the growth index and the position of the peaks in the CMBR anisotropy spectrum. It is, therefore, interesting to apply the method outlined above to find out the $f(R)$ counterpart of this model. To this aim, we just need to insert the expressions for $a(t)$ and $H(t)$ into Eqs.(\ref{eq: singleeq})\,-\,\ref{eq: in1bis}) and numerically determine first $f(t)$ and then $f(R)$. As a preliminary step, we set ${\cal{H}}_0 = 0.97$ as obtained in \cite{ester}. To estimate $f''(R_0)$ through Eq.(\ref{eq: in2}), we use $t_0 = 13.7 \ {\rm Gyr}$ \cite{WMAP} and $(\dot{G_{eff}}/G_{eff})_{obs} = 1.6 {\times} 10^{-12} \ {\rm yr^{-1}}$ that is the upper limit determined through heliosismological studies \cite{helio}. Note that this choice is somewhat arbitrary since there are a lot of different estimates obtained with radically different methods \cite{uzan}. For instance, using Lunar Laser Ranging, Williams et al. \cite{LLR} have recently obtained $\dot{G}_{eff}/G_{eff} = (4 \pm 9)\times 10^{13} \ {\rm yr^{-1}}$. Fortunately, most of the estimates available in literature are all consistent within the uncertainties so that we are confident that any bias is not introduced in the reconstruction process.

\begin{figure}
\centering
\resizebox{8.5cm}{!}{\includegraphics{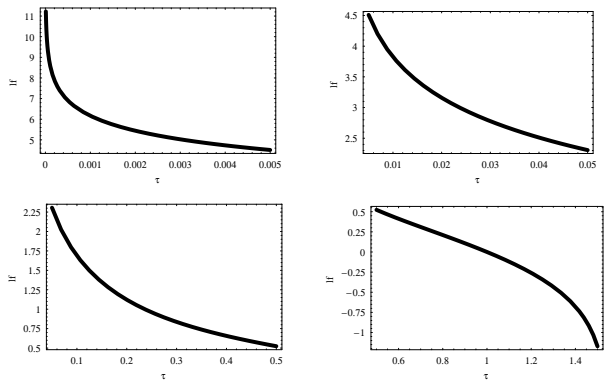}}
\caption{Reconstructed $f(t)$ as function of $\tau = t/t_0$. For clarity, we plot $lf \equiv \log{(f/f_0)}$ rather than $f$ itself. The upper left panel refers to the period near the recombination epoch, while the lower right panel extends in the near future.}
\label{fig: ftsingle}
\end{figure}

Before discussing the results, there is a conceptual point to clarify. The method described in Sect.\,III represents a sort of {\it bridge} between two different scenarios. In particular, their matter contents could be different, so that we should define both $\Omega_M^{curv}$ and $\Omega_M^{sf}$ to denote this quantity in the two different models. In principle, there is no reason why $\Omega_M^{curv} = \Omega_M^{sf}$ should hold. However, since $\Omega_M^{sf}$ is close to the fiducial value ($\Omega_M \sim 0.3$) suggested by model independent estimates (e.g., from galaxy clusters abundance), we take $\Omega_M^{curv} = \Omega_M^{sf}$.

The reconstructed $f(t)$ for the exponential potential model is shown in Fig.\,\ref{fig: ftsingle}, where we plot $\log{(f/f_0)}$ as function of the dimensionless time $\tau$. It is, however, more interesting to look at Fig.\,\ref{fig: frsingle}, where $\eta \equiv (f/f_0)/(R/R_0)$ is plotted in terms of the scalar curvature $R$ normalized to its present day value $R_0$. Although an analytical solution has not been obtained, we have found that the numerically reconstructed Lagrangian may be very well approximated\footnote{Defining $\epsilon \equiv \left [ f(R) - f_{app}(R) \right ]/f(R)$, with $f$ and $f_{app}$ the numerically reconstructed and the approximating $f(R)$, we get $| \epsilon | \le 0.06$ over the full time range $10^{-5} \le \tau \le 1$, which is indeed quite satisfactory.} as\,:

\begin{figure}
\centering \resizebox{8.5cm}{!}{\includegraphics{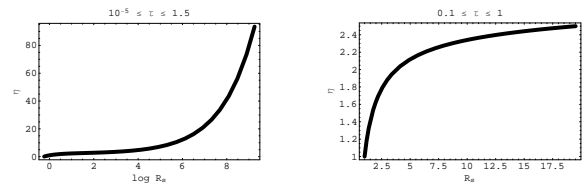}}
\caption{Reconstructed $f(R)$ for the exponential potential model. In both panels, $\eta \equiv (f/f_0)/(R/R_0)$ is plotted for sake of clarity. Left panel shows the result of the reconstruction process over the full time range $10^{-5} \le \tau \le 1.5$, so that we report $\log{R_s}$ (with $R_s \equiv R/R_0$) on the abscissa for sake of clarity. In the right panel, we zoom on the epoch $0.1 \le \tau \le 1$ corresponding to $0 \le z \le 4.7$. Note that $R$ is a decreasing function of $t$ so that the time increases towards right.}
\label{fig: frsingle}
\end{figure}

\begin{eqnarray}
f(R) & \simeq & \eta_s f_0 \left ( \frac{R}{R_0} \right ) \nonumber \\
~ & ~ & ~ \\
 ~ & \times &  \left \{1 + \left [ \eta_1
\left ( \frac{R}{R_0} \right )^{\alpha} + \eta_2 \left ( \frac{R}{R_0} \right )^{-\beta} \right ] \ln{\left ( \frac{R}{R_0} \right )} \right \} \ , \nonumber
\label{eq: frfit}
\end{eqnarray}
where the parameters $(\eta_s, \eta_1, \eta_2, \alpha, \beta)$ depend on the value adopted for $f''(R_0)$. For our case, it is\,:

\begin{displaymath}
(\eta_s, \eta_1, \eta_2, \alpha, \beta) = (1.07, 0.029, 0.810, 0.232, 0.221) \ .
\end{displaymath}
The approximating function in Eq.(33) indeed shows to be a useful tool to discuss the main features of the reconstructed $f(R)$.

First, let us note that, for $R >> R_0$ (that is, for very small $\tau$), the second term in the square brackets may be neglected and $f(R) \sim R + \eta_1 R^{1.23} \ln{R}$. This model can be particularly interesting in the early universe, where it allows inflation \cite{Star}. In our case, however, this regime is achieved also during the recombination epoch, when nucleosynthesis took place. In this limit, one should expect that standard general relativity (i.e., $f(R) \propto R$) is recovered in order to not alter the successful theory of primordial nucleosynthesis. Actually, the formation of light elements depends on the physics of nuclear reactions that, of course, is independent of the form of $f(R)$ and on the expansion rate $H(t)$ setting the beginning and the end of the nucleosynthesis process. This latter ingredient is the same in our model as in the standard recipe, since $f(R)$ has been reconstructed in such a way that $H(t)$ has the same functional dependence on $t$ as in the model with a scalar field and no departures from general relativity. As such, we conclude that having a nonstandard $f(R)$ during the recombination epoch does not affect the process of formation of primordial elements.

As the time increases, the $f(R)$ never approaches a linear function, so that the general relativity Lagrangian $f(R) \sim R$ is never approached. In particular, for $0.05 \le \tau \le 0.5$ (i.e., $1 \le z \le 8$), both terms in Eq.(33) contribute to the Lagrangian giving rise to large departures from the standard model. Considering that the typical galaxy formation redshift is of order $z_F \sim 2 \div 5$, it could thus be interesting to investigate whether this could have any impact on the theory of galaxy formation.

For $0.5 \le \tau \le 1$ (i.e., for $z \le 0.8$, part of the redshift range probed by the SNeIa Hubble diagram), the best fitting function significantly differs from standard general relativity as expected. Moreover, we note that $f(R)$ is neither a simple power law in $R$ (or a combination of power law functions), nor a logarithmic one. This could be surprising since both these classes of Lagrangians have been widely discussed in literature and found to be in agreement with the data. Actually, it is well known that the data do not select a unique class of models (dark energy\,-\,like or higher order gravity theories). Moreover, the reconstructed $f(R)$ predicts the same dynamics of the scalar field model with the exponential potential, so that we are sure that it fits the data equally well. However, it is intriguing to note that the approximating function is a combination of both (power law and logarithmic) classes of $f(R)$ theories.

\section{Conclusions}

Summarizing, we may conclude that the $f(R)$ theory reconstructed starting from the exponential potential is able to solve the puzzle of cosmic acceleration resorting to corrections to general relativity that are significative both in the early and late universe. As we have discussed, however, considering $f(R) \propto R$ during the nucleosynthesis epoch is not mandatory to get the correct abundance of light nuclei, so that this argument may not be used to reject the reconstructed $f(R)$'s.

On the other hand, the impact of strong departures from general relativity during the galaxy formation epoch should be investigated with much care. Actually, while the time scale of the process is unaltered since the expansion rate of the universe is the same, the physics of the gravitational collapse giving rise to the protogalactic clouds may be different in a fourth order theory with respect to quintessence models (independent of the shape of the potential). Indeed, the Newtonian potential obtained considering the weak field limit of a particular class of these theories (namely $f(R) \propto R^n$) is modified and a corrective term as $r^\alpha$ has to be added to the $1/r$ Newtonian term \cite{newtlimitnoi}. However, it is worth noting that it is indeed possible that the strength of the correction is too low to significantly alter the process, so that any difference could be hardly detected.

There is a subtle point to be discussed inherently to the the peculiar nature of the reconstruction procedure we have employed. As explained in Sect.\,III, we have set boundary conditions at the present time and integrated Eq.(\ref{eq: singleeq}) backward in time. As an alternative procedure, one could give boundary conditions at some very far point in the past and then integrate forward. Indeed, a possible strategy could be to impose that $f(R_{in}) \sim R$ (and hence $f'(R_{in}) = 1$, $f''(R_{in}) = 0$) with $R_{in} = R(\tau_{in})$ and $\tau_{in} \sim 10^{-5}$, that is, to impose that standard general relativity is recovered during the recombination epoch. Performing such a reconstruction, we infact get a $f(R)$ function that departs from the general relativistic one ($f(R) \sim R$) only in the recent epochs, so that all the successful results of the standard theory are automatically retained. Although such a feature is highly desirable, this approach is manifestly biased by theoretical prejudices. After all, there is no aprioristic reason why $f(R)$ should be linear in $R$ on scales that have never been directly probed by observations.

It is also interesting to stress that the reconstructed $f(R)$ is by no way unique since different boundary conditions lead to diverse $f(R)$ theories, all sharing the feature of giving rise to the same expressions for the Hubble parameter and the scale factor. This result does not come as a surprise. For each $f(R)$, there is a different Lagrangian giving rise to a particular set of field equations. Since these latter are nonlinear, there may be, in general, more than one solution, and it is thus not unlikely that a given solution is common to more than one $f(R)$ theory. Going backward from solutions to $f(R)$ therefore cannot uniquely determine $f(R)$, and this explains why using different boundary conditions leads to different higher order theories. We stress, however, that this cannot be considered as a weakness of our reconstruction procedure. Our aim is, in fact, to find out a class of $f(R)$ theories able to reproduce a given cosmological dynamics, and this is indeed what we do successfully. Choosing among the possible theories can only be done by carefully investigating what are the most suitable boundary conditions. This is outside the aim of the present work, but will be discussed in a forthcoming paper.

While the reconstructed $f(R)$ correctly works on cosmological scales, its consequences on the Solar system dynamics are worth investigating. Actually, this is an open problem for all the class of fourth order theories and, up to now, no definitive conclusions have been drawn. A possible way out is to evaluate PPN parameters for the case of $f(R)$ gravity exploiting the analogy between fourth order theories and scalar\,-\,tensor ones. Such an analysis has been indeed performed by some of us \cite{CT} and will be presented in a forthcoming work.

As a concluding remark, we would like to emphasize the resulting substantial equivalence between two radically different scenarios, namely scalar field quintessence and fourth order theory of gravity. Since it is possible to find $f(R)$ in such a way that the same dynamics is predicted by the two models, infact, we are in presence of an intrinsic degeneracy that cannot be broken by observations probing only dynamical quantities (but see the discussion in the concluding section of \cite{CCT}). While this is bad news for observational cosmology, it is a {\it call to army} for theoreticians to look for suitable and self consistent motivations to choose between these two equivalent approaches to the dark energy puzzle.

\acknowledgments{We warmly thank Paolo Scudellaro, Crescenzo Tortora and Antonio Troisi for the discussions and a careful reading of the manuscritpt.}

\end{document}